\documentclass{ws-p8-50x6-00}

\begin{document}

\title{Heavy quark production in the semihard QCD approach
at HERA and beyond}

\author{S.P. Baranov}

\address{P.N. Lebedev Physics Institute, Leninsky prosp. 53,  
Moscow 117924, Russia\\ 
E-mail: baranov@sci.lebedev.ru}

\author{A.V. Lipatov}

\address{Department of Physics, M.V. Lomonosov Moscow State
University, Moscow 119899, Russia\\
E-mail: artem\underline{\hphantom{3}}lipatov@mail.ru}

\author{N.P. Zotov}

\address{D.V. Skobeltsyn Institute of Nuclear Physics,
 M.V. Lomonosov Moscow State University,
 Moscow 119899, Russia\\  
E-mail: zotov@theory.sinp.msu.ru}


\maketitle

\abstracts{
Processes of heavy quark production at HERA, TEVATRON and THERA
energies are considered using the semihard ($k_T$ factorization)
QCD approach with emphasis on the BFKL dynamics of gluon
distributions.}

The experimental results on $b\bar b -$pair production cross sections
obtained by the H1 and ZEUS Collaborations at HERA\cite{H}
 and D0 and CDF Collaborations at TEVATRON\cite{T}
 provide a strong impetus for further
theoretical studies.  Comparisons of these results with NLO
pQCD calculations show that they underestimate the cross sections
at HERA and TEVATRON energies. Therefore, it looks certainly
reasonable to try a different approach.

 In this work we focus on the description of $b\bar b-$pair cross sections
at HERA and TEVATRON in the so called semihard ($k_T$ factorization) QCD
approach (SHA)\cite{GLR,CCH},
 which we have applied earlier to open charm\cite{SZ} and
 $J/\Psi$ photoproduction at HERA (see in ref.\cite{SZ}). We also discuss 
the sensitivity of our theoretical results\cite{BZ3} to the BFKL
 type dynamics\cite{BFKL} which may be investigated in the photoproduction
of $D^{*}$ and $J/\Psi$ mesons at THERA energies. 

 In SHA, the unintegrated gluon distribution $\varphi_G(x, k_{T}^2)$
is connected with the conventional 
gluon density $xG(x, Q^2)$ by the following relation
\begin{equation}   
 xG(x, Q^2) =  xG(x, Q_0^2) + \int_{Q_0^2}^{Q^2} dk_T^2 
\varphi_G(x, k_{T}^2), 
\end{equation}
where $Q_0^2$ is the collinear cutoff parameter.
\begin{figure*}[ht]
\begin{center}
\includegraphics[width=0.45\linewidth]{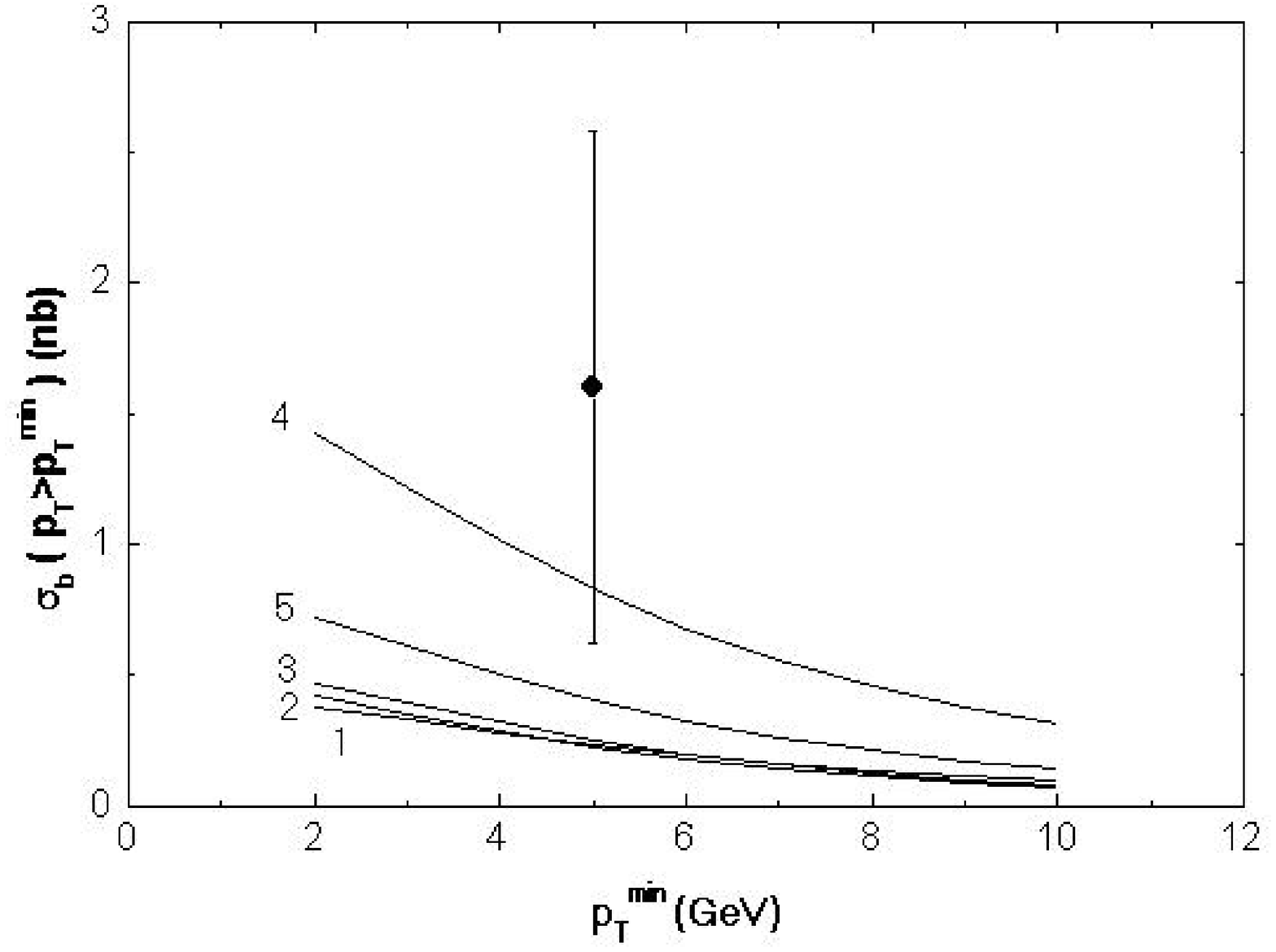}
\includegraphics[width=0.48\linewidth]{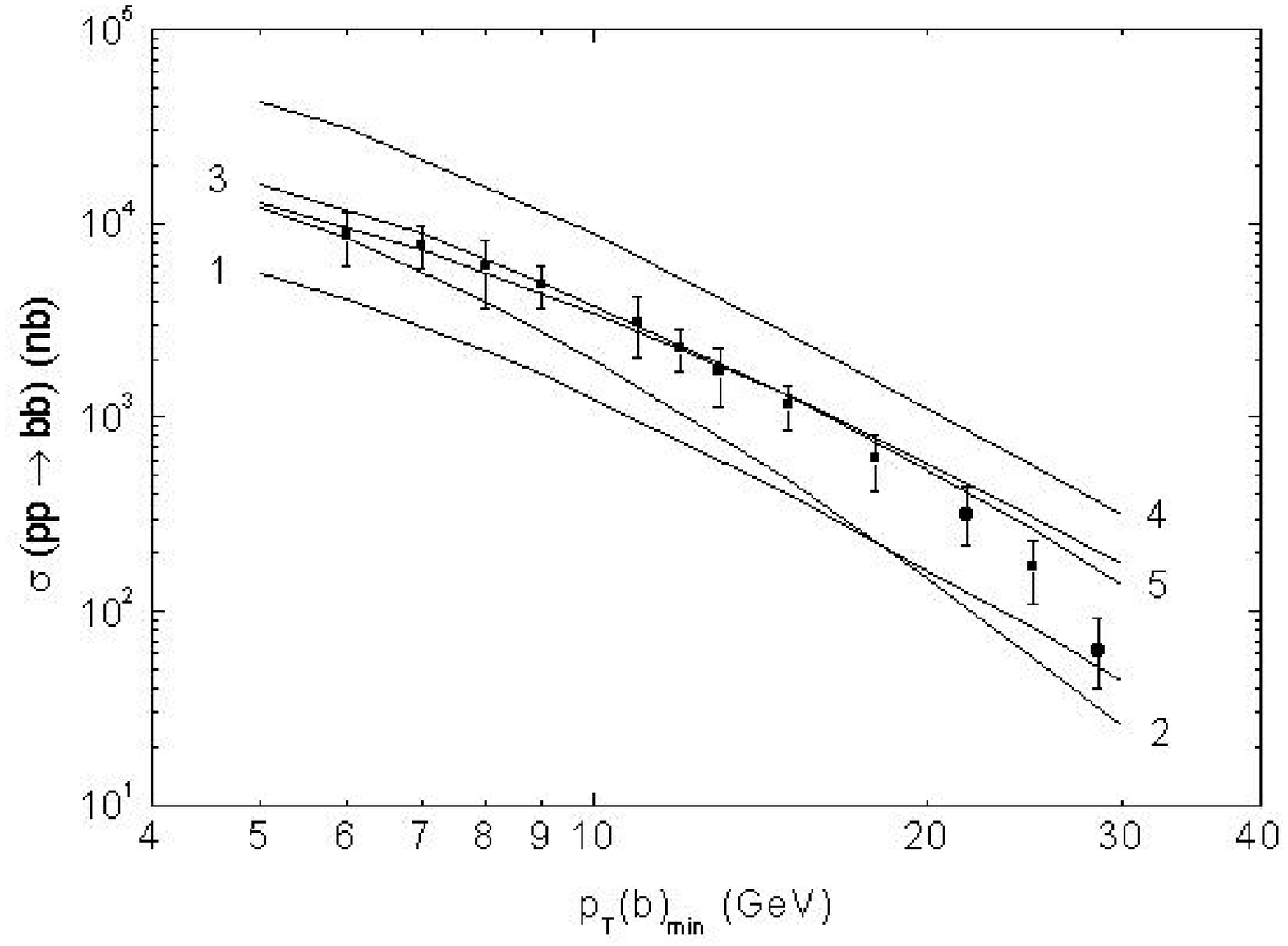}
\end{center}
\vspace*{-5mm}
\caption{The cross sections of $b\bar b$ production $\sigma(p_T > p_T^{min})$
 at HERA (left panel) and  TEVATRON (right panel): curves 1, 2, 3, 4 and 5 
correspond to the
MT, GRV, RS, LRSS and BFKL parametrizations of gluon distribition.}
\end{figure*}
  We used the results of ref.\cite{CCH}  for the off mass shell parton 
cross sections, and we used several different parameterizations for the
unintegrated gluon distribution (see ref.\cite{SZ} for details),
namely: the LRSS\cite{GLR},
RS\cite{RS} and the so called BFKL~\cite{Blum} parameterizations.
We used the following set of SHA parameters: $Q_0^2 =$4, 2 and 1 GeV$^2$ in (1)
for the RS, LRSS  and BFKL parameterizations; in the case of the BFKL
parameterization the parameter $\Delta = 0.35$\cite{SZ}; everywhere the
charm and beauty  quark masses are $m_c =$1.5 GeV and $m_b =$4.75 GeV.

 The results of our calculations for the total cross section of inelastic
$b\bar b$ photoproduction at HERA as compared to H1\cite{H} data are
\begin{figure*}[ht]
\begin{center}
\includegraphics[width=0.45\linewidth]{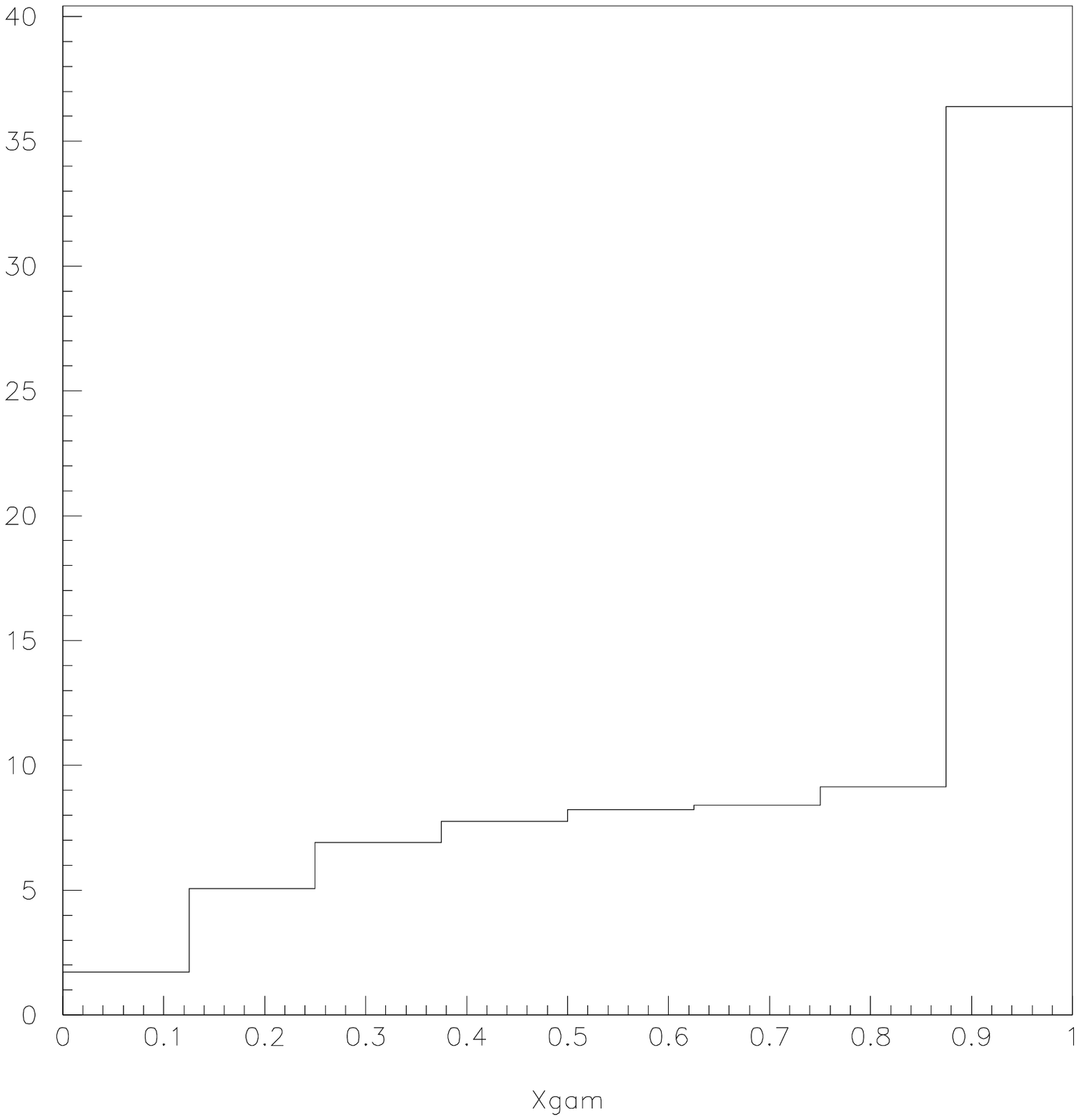}
\includegraphics[width=0.45\linewidth]{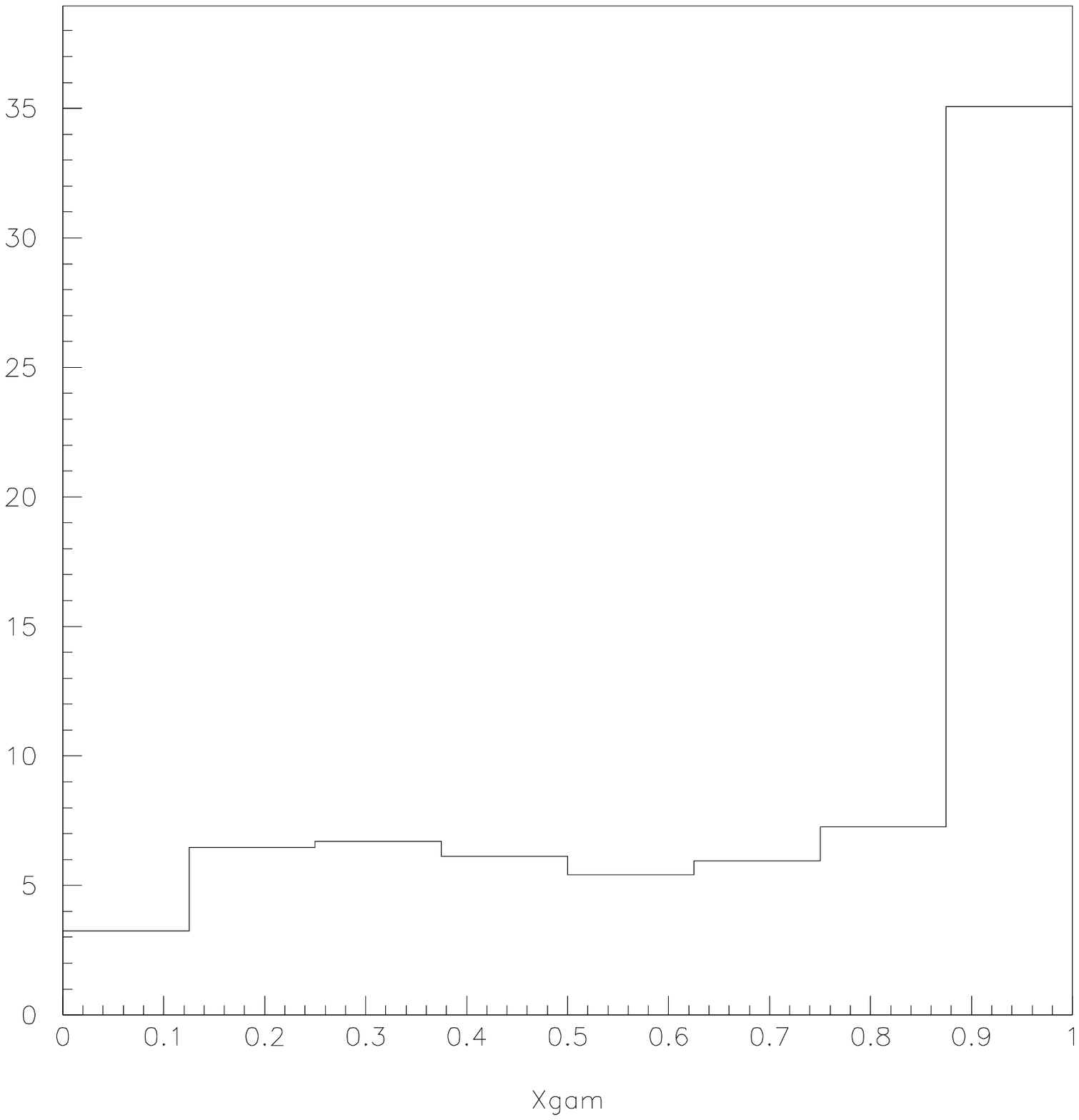}
\end{center}
\vspace*{-5mm}
\caption{The differential cross section $d \sigma /dx_\gamma$ (nb) for 
$Q^2<1$ GeV$^2$ with BFKL (left panel) and  CCFM (right panel)
 unintegrated gluon distributions
at THERA.}
 \label{xgamcc}
\end{figure*}
\begin{figure}[!t]
\begin{center}
\epsfig{figure=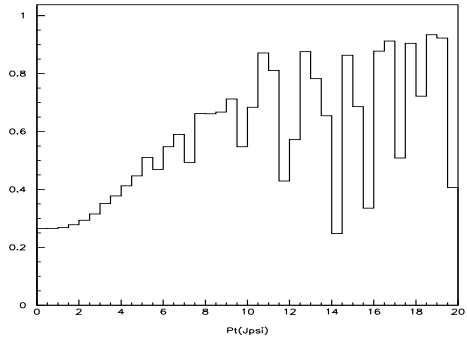,width=7cm,height=5cm}
\end{center}
\caption{The fraction of $J/\Psi$ mesons in helicity
zero state (degree of spin alignment).}
\label{fig3}
\end{figure}
published in the paper by Lipatov, Saleev, Zotov\cite{SZ}. We have shown
there that the H1 data are well described by the LRSS parametrization
and by the the BFKL parametrization but only with as small $m_b$ as 
$m_b =$4.25 GeV in the latter case. In Fig. 1a we show our
results for the total cross
section of inelastic $b\bar b$ photoproduction at HERA compared to
 ZEUS data\cite{H}.
We see that only the LRSS parametrization describes the ZEUS data 
(at $m_b =$4.75 GeV). In contrast with this, the cross
section for $b\bar b$
production at TEVATRON\cite{T} is described by the BFKL and RS
parametrizations very well (Fig. 1b). The LRSS parametrization
(at the same values\cite{SZ} of parameters and normalization)
overshoots the D0 (and CDF) data.

 In the ref.\cite{BZ2} the calculations of the associated charm
and dijet production cross section have been made within the SHA
with BFKL and CCFM\cite{HJ} unintegrated gluon distributions at HERA
energies. 
The attention was focused there on the variable $x_\gamma$, which is the
fraction of the photon momentum contributed to a pair of jets with
largest $p_T$. The results of the similar calculations made for THERA
conditions  are shown in Fig. 2  as a futher test of the underlying dynamics.
 The existence of the wide plateau at $x_\gamma < 0.9$ seen in the fugure
comes from the noncollinear gluon evolution, which generates gluons with
non-negligeable transverse momentum. In a significant fraction of events 
the gluon emitted close to the quark box appears to be even harder than
one or even both of the quarks produced in hard interaction.

The effects of initial gluon off-shellness may be, best of all, 
seen in the transverse momentum spectra of $J/\Psi$ mesons\cite{B}.
 In contrast with the conventional (massless) parton model, the SHA shows
that the fraction of $J/\Psi$  mesons in the helicity zero state increases
with their transverse momentum $p_T$. A deviation from the parton model
behaviour becomes well pronounced already from $p_T > 3$ GeV at HERA  
energies\cite{B}, and at $p_T > 6$ GeV the helicity zero polarization
tends to be dominant. The same effect is seen in Fig. 3, where
we show the results of the calculations\cite{BZ3} of the ratio 
$\sigma_{h= 0}/\sigma$ for $J/\Psi$ photoproduction at THERA conditions
made with the BFKL unintegrated gluon distribution. 

The examples considered in this paper demonstrate the effects of the BFKL
gluon evolution on the important and experimentally measurable quantities,
such as the event topology or vector meson spin alignement. At present, 
the theoretical predictions made for HERA and TEVATRON 
conditions have found their
experimental confirmation. A further investigation of the relevant effects
at THERA collider can put our understanding of the hadron structure on
even more solid grounds.

One of us (N.Z.) is gratefull  to Russian Academy of Science and
 the Organizing Committee of  DIS2001 for financial support.
This work has been supported by the Royal Swedish Academy of Sciences.

\end{document}